\documentclass[a4paper,11pt]{article}

\usepackage{pos}
\usepackage{subfig} 
\usepackage[export]{adjustbox} 
\usepackage{float} 
\usepackage{multirow} 

\title{Development of a trigger for acoustic neutrino candidates in KM3NeT}

\author[a,1]{Miguel\ Ardid}
\author[b,1]{Manuel\ Bou-Cabo}
\author*[a,1]{Dídac\ D.Tortosa}
\author[b,1]{Guillermo\ Lara}
\author[a,1]{Juan A.\ Martínez-Mora}

\affiliation[a]{Institut d’Investigació per a la Gestió Integrada de les Zones Costaneres (IGIC), Universitat Politècnica de València (UPV),\\
Paranimf 1, 46730 Platja de Gandia, València}

\affiliation[b]{Instituto Español de Oceanografía (IEO), Mixed unit UPV-IEO,\\
Tinglados Muelle frutero, 46730 Grau de Gandia, València}

\note{For the KM3NeT collaboration.}

\emailAdd{mardid@fis.upv.es}
\emailAdd{manuel.bou@ieo.es}
\emailAdd{didieit@upv.es}
\emailAdd{guillermo.lara@ieo.es}
\emailAdd{jmmora@fis.upv.es}

\abstract{The KM3NeT Collaboration is constructing two large neutrino detectors in the Mediterranean Sea: ARCA, located near Sicily and aiming at neutrino astronomy, and ORCA located near Toulon and designed for the study of intrinsic neutrino properties (neutrino oscillations, mass ordering, etc.). The two detectors together will have hundreds of Detection Units (DUs) with Digital Optical Modules (DOMs) kept vertically by buoyancy forming a large 3D optical array for detecting the Cherenkov light produced after the neutrino interactions. To properly reconstruct the direction of the incoming neutrino, the position of the DOMs, which are not static due to the sea currents, must be monitored. For this purpose, the detector is equipped with an Acoustic Positioning System, which is composed of fixed acoustic emitters on the sea bottom, a hydrophone in each DU base, and a piezoceramic sensor in each DOM, as acoustic receivers. This network of acoustic sensors can be used not only for positioning, but also for acoustic monitoring studies such as bioacoustics, ship noise monitoring, environmental noise control, and acoustic neutrinos detection. This work explores the possibility of creating a trigger for saving the data for ultra-high-energy neutrino candidates detected acoustically by the hydrophones. The acoustic signal caused by the neutrino interaction in a fluid is a short-time duration Bipolar Pulse (BP) extremely directive and with a Fourier transform extending over a wide range of frequencies. A study of signal detection, out of the background, has been done by simulating BP produced by the interaction of a 10$^{20}$ eV neutrino at 1 km from the detector at zero-degree incidence added to the experimental real acoustic data. Finally, a trigger proposal has been developed in order to record candidates of BPs and it has been tested. The number of candidates per second, precision, and recall have been monitored according to the cuts applied and parameters calculated by the algorithm.}

\FullConference{%
  9th International Workshop on Acoustic and Radio EeV Neutrino Detection Activities - ARENA2022\\
  7-10 June 2022\\
  Santiago de Compostela, Spain}

\renewcommand{\logo}{\relax}

\begin{document}
\maketitle

\section{KM3NeT}
The Cubic Kilometre Neutrino Telescope (KM3NeT) is an underwater detector in the Mediterranean Sea. It is currently under construction and is intended to be the largest deep-water telescope in the world. This type of telescope detects the Cherenkov light that the interaction of a high-energy muon neutrino may produce in the fluid (water or ice).  For this purpose, it installs a 3D network of Digital Optical Modules (DOMs) spread over a huge volume of water. A Detection Unit (DU) is composed of a base anchored to the seabed, 18 DOMs attached to it by a pair of cables and maintained vertically by their own buoyancy and that of a top buoy \cite{KM3NeT2016}. 

\subsection{The detector framework}
KM3NeT has two detector nodes called Astroparticle Research with Cosmics in the Abyss (ARCA) and Oscillation Research with Cosmics in the Abyss (ORCA). Both use the same technology, although for different purposes: ARCA is designed to study very high energy cosmic neutrinos (in the TeV-PeV range) \cite{KM3NeT2019-9}, while ORCA studies the properties of atmospheric neutrinos in the 3-100 GeV range \cite{KM3NeT2020-10}. ARCA is located 100 km off the coast of \textit{Portopalo di Capo Passero} at 3400 m depth, and ORCA 40 km off the coast of \textit{Toulon} submerged at 2400 m.

	
ARCA will occupy about 1 km$^3$, distributing a total of 220 DUs in two blocks. ORCA would occupy a volume of 0.018 km$^3$ in a space appropriated by 120 DUs in a single block. The main difference between ARCA and ORCA is the height of their DUs (700 m at ARCA and 190 m at ORCA) and the distance between them. The first DU in ARCA was installed in December 2015 and currently has 21 DUs deployed and operational, while the first DU in ORCA was installed in February 2019 and so far has 11 operational DUs.

\subsection{Acoustic sensors}
\begin{figure}[htbp]
		\centering
		\includegraphics[width=12cm]{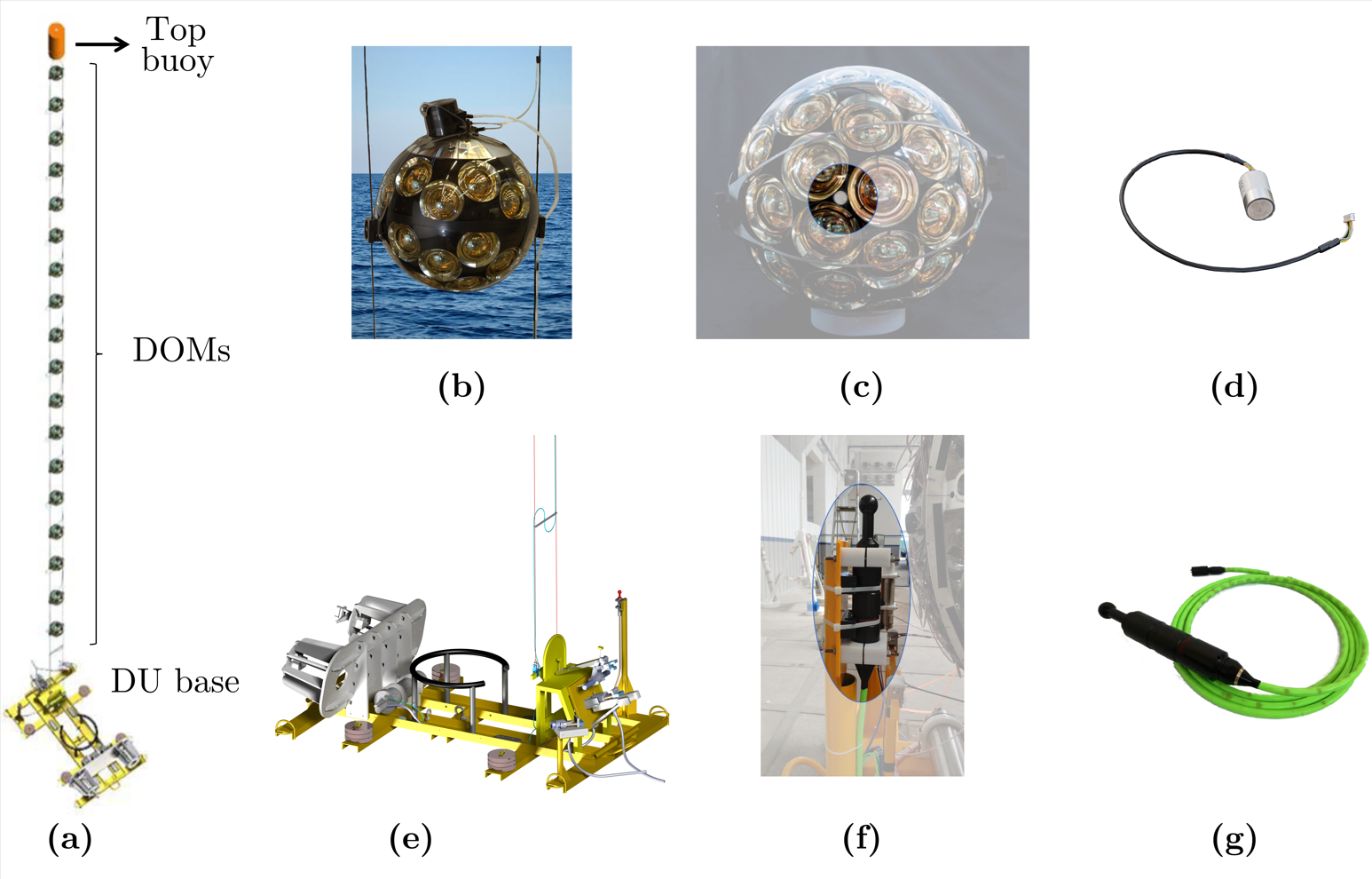}
		\caption{\textbf{(a)} A DU scheme. \textbf{(b)} A DOM. \textbf{(c)} Piezoceramic sensor in DOM. \textbf{(d)} Piezoceramic sensor. \textbf{(e)} DU base sketch with hydrophone. \textbf{(f)} Hydrophone on a DU base. \textbf{(g)} Hydrophone.}
		\label{fig:Sensors}
	\end{figure}

Deep sea currents prevent the DUs from remaining fully vertical. Since the position of these sensors should be known with good accuracy ($\sim$10 cm) in order to reconstruct the interacting neutrino direction with a good angular resolution, KM3NeT is equipped with an Acoustic Positioning System (APS). The APS makes use of acoustic sensors: the emitters, anchored to the seabed in a known position, and the receivers, a hydrophone at the base of each DU and a piezoceramic sensor in each DOM (see \textit{\autoref{fig:Sensors}}).

\section{Acoustic neutrino detection}
\subsection{Bipolar Pulse}
\begin{figure}[htbp]
	\centering
	\subfloat[]{\includegraphics[height=4.5cm]{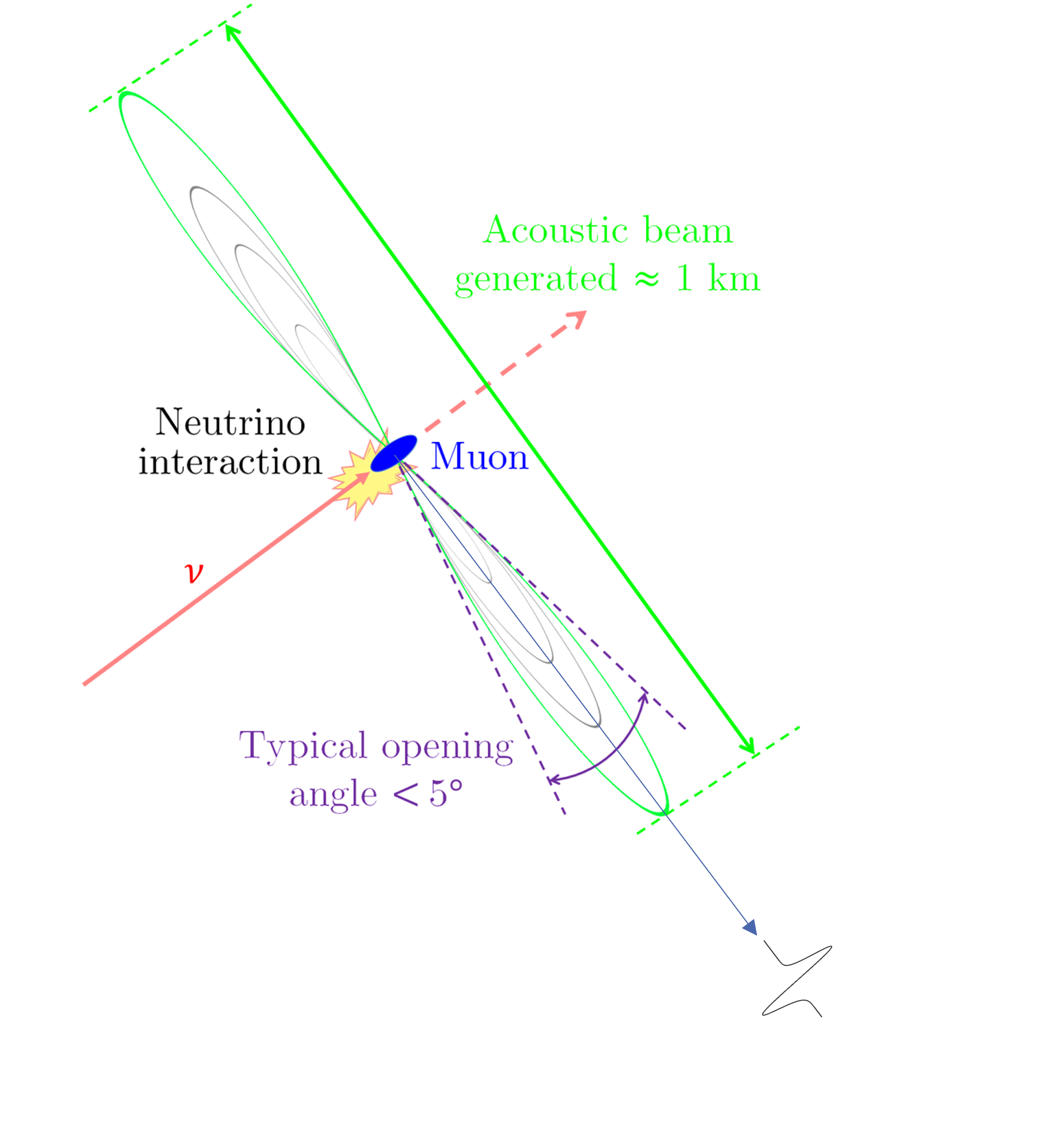}}
	\hspace{0.7cm}
	\subfloat[]{\includegraphics[height=4.5cm]{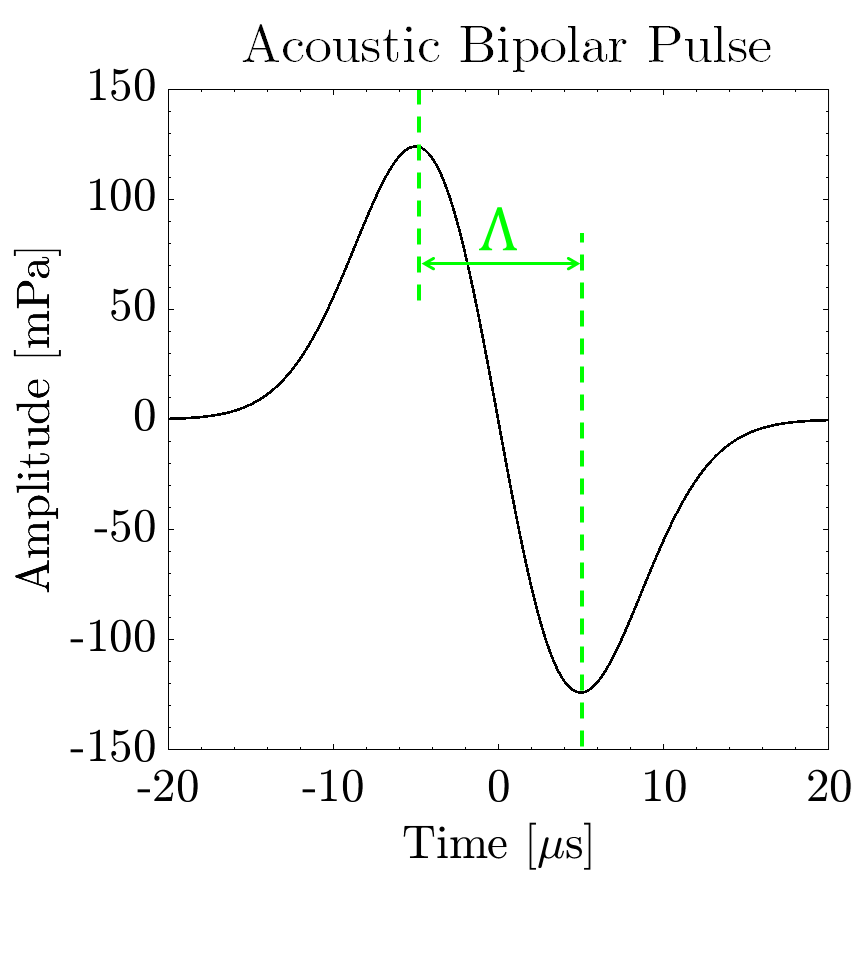}}
	\hspace{0.7cm}
	\subfloat[]{\includegraphics[height=4.5cm]{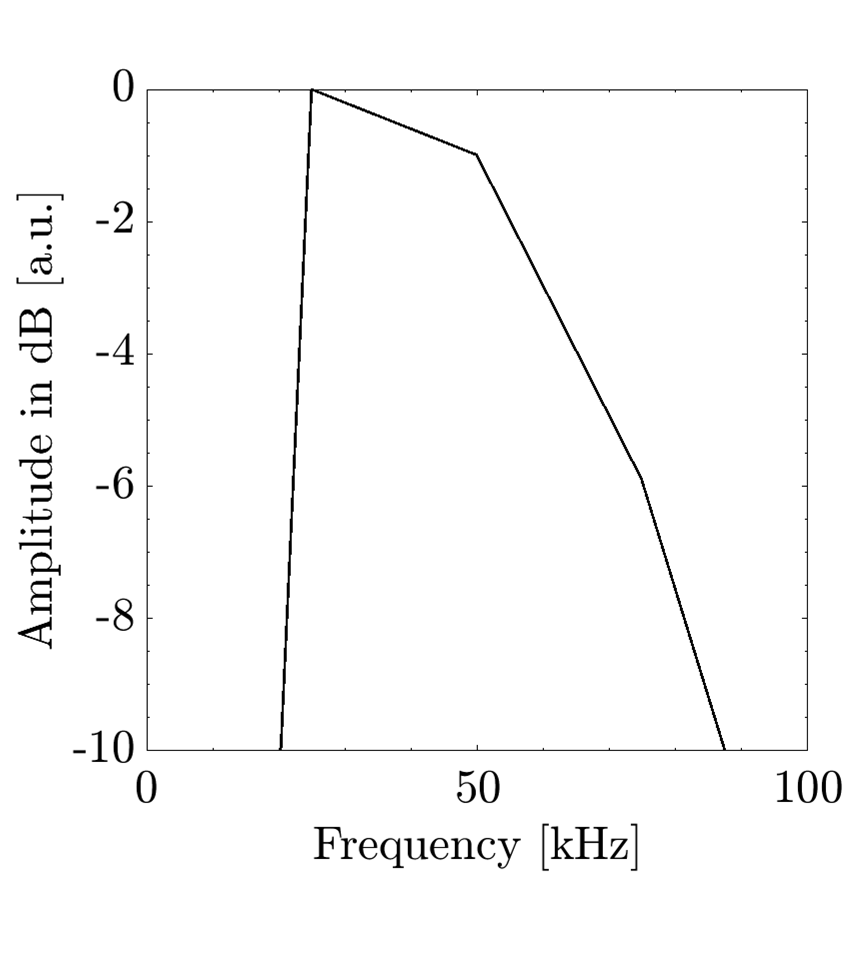}}
	\caption{\textbf{(a)} Acoustic "pancake" beam generation from a neutrino interaction. \textbf{(b)} Acoustic BP simulating the signal after neutrino interaction on a fluid. The amplitude is from a neutrino deposition of 10$^{11}$ GeV of thermal energy at 1 km and 0 degrees \cite{ACORNE2009}. \textbf{(c)} Acoustic BP in frequency domain ($f_s=$ 20 MHz).}
	\label{fig:NeutrinoAcousticSign}
\end{figure}

When Ultra-High-Energy (UHE) neutrino, $\ge1$ EeV, interacts in water generates a thermo-acoustic pulse. This short Bipolar Pulse (BP) propagates with a very narrow beam directivity (less than 5 degrees) and contains a wide range of frequencies below 100 kHz (see \textit{\autoref{fig:NeutrinoAcousticSign}}) \cite{Askariyan1979}. 

\begin{figure}[htbp]
        \centering
        \subfloat[]{\includegraphics[height=4cm]{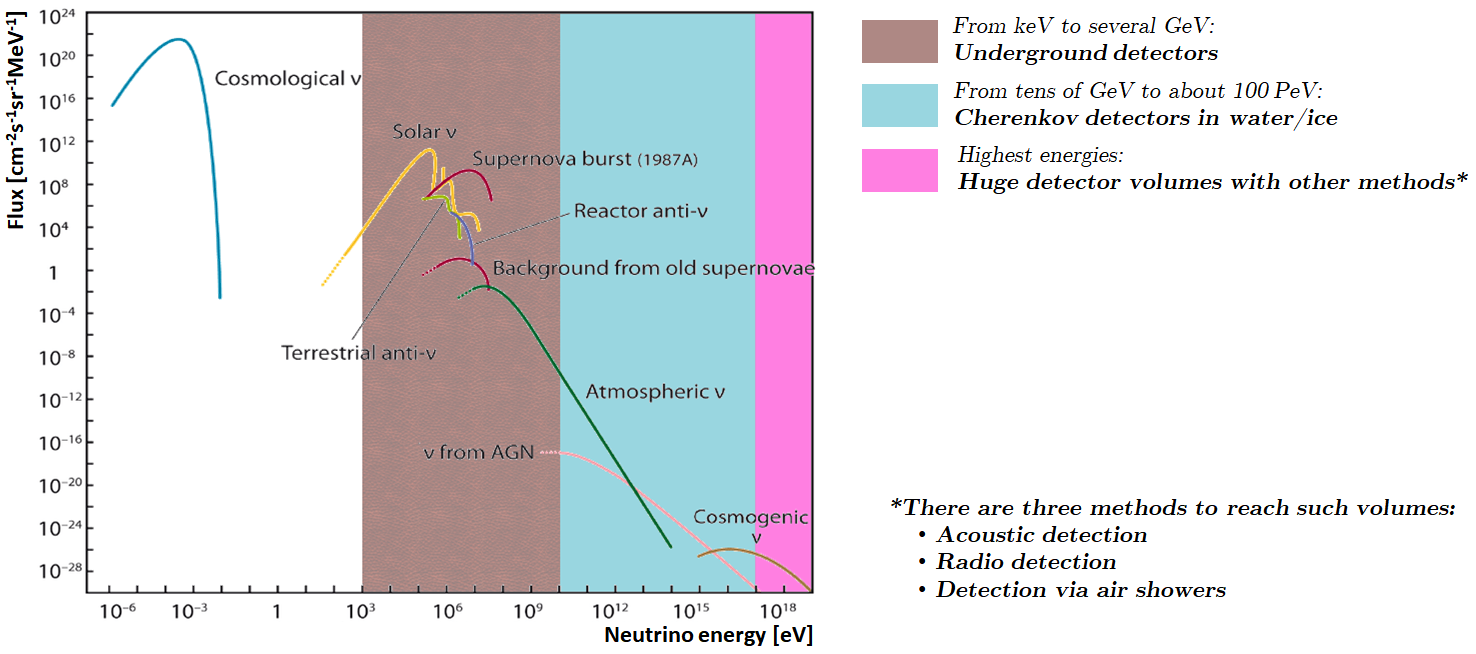}}
        \hspace{0.3cm}
        \subfloat[]{\includegraphics[height=3.9cm]{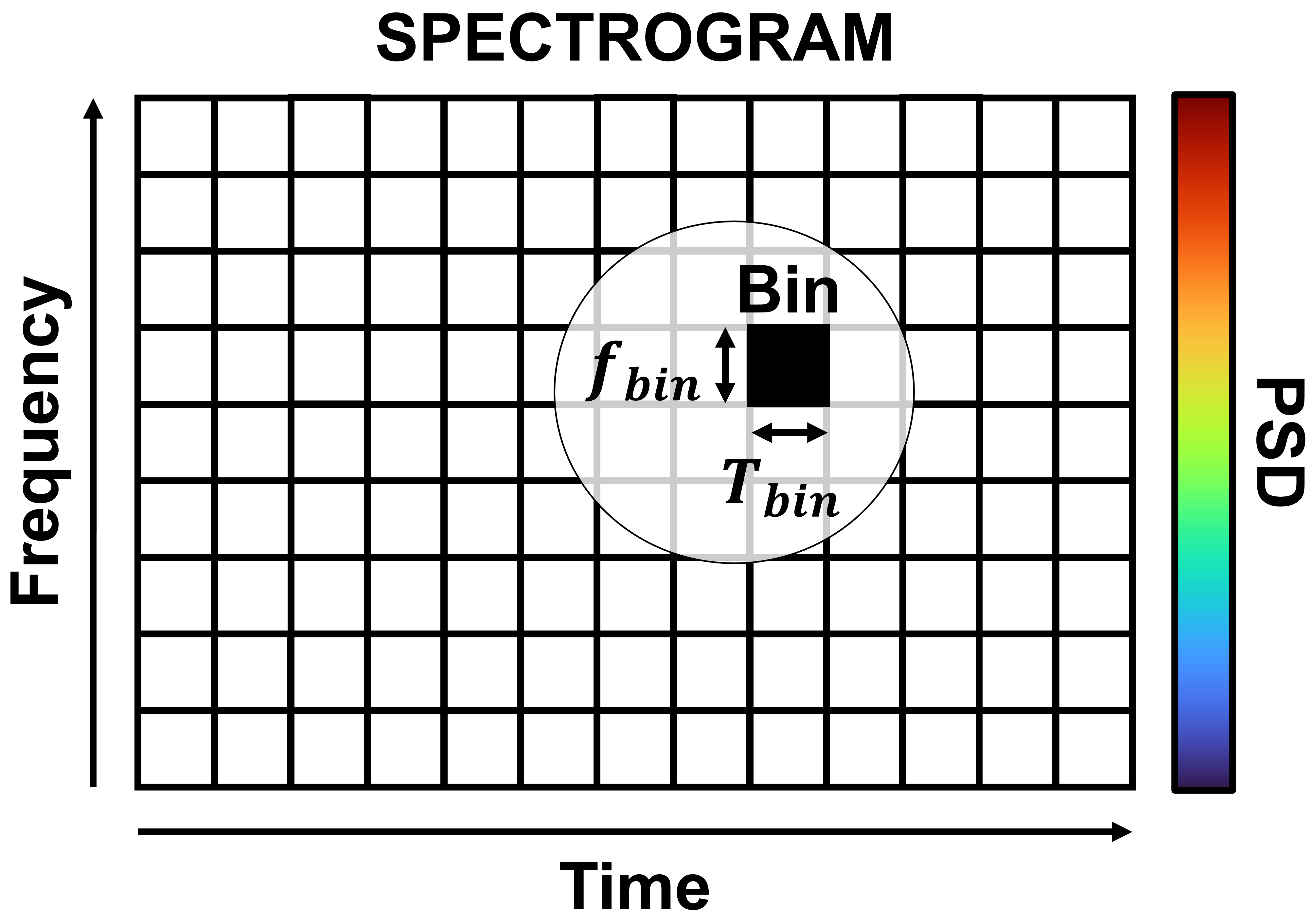}}
        \caption{\textbf{(a)} Measured and expected fluxes of natural and reactor neutrinos \cite{Spiering2012}. \textbf{(b)} Sketch of spectrogram.}
        \label{fig:AcouDetect}
    \end{figure}

There are studies that estimate the possibility of detecting UHE neutrinos with other types of technology than optics (acoustic, radio, or via air showers detection) since it is a huge volume to be monitored (see \textit{\autoref{fig:AcouDetect}}\textit{\color{blue}{.a}}\color{black}) \cite{Spiering2012}.
Ultrasonic signals in water have lower attenuation compared to light, favouring acoustic detection. Although the difficulty of capturing one of these signals lies not only in the wide frequency component that makes its identification difficult, or its narrow directivity, but also in its short duration (the distance between BP peaks is estimated to be of the order of 10 $\mu$s), which requires a system with a sampling rate ($f_s$) capable of capturing the shape of the BP.

\subsection{The spectrogram as a signal detector}
The spectrogram represents in three dimensions the time, the frequency, and amplitude of the energy distribution of a signal. The X--axis represents the time, the Y--axis the frequency, and the Power Spectral Density (PSD) is coloured coded. A spectrogram decomposes the signal into bins. Each bin represents the PSD value in a frequency-temporal interval (see \textit{\autoref{fig:AcouDetect}}\textit{\color{blue}{.b}}\color{black}). A bin is created using a given number of signal samples, $N_{bin}$, and will represent the PSD value contained in it. 
It should be noted that the time and frequency resolution, $T_{bin}$ and $f_{bin}$ respectively, of a spectrogram is limited by the pre-selected $N_{bin}$ and the $f_s$ of the recorded signal:
\begin{equation}\label{eq:_Tbin}
	T_{bin}\ge \frac{N_{bin}}{f_s} \longrightarrow N_{bin}\le T_{bin}\cdot f_s
\end{equation}
\begin{equation}\label{eq:f_res}
	f_{bin}\ge \frac{f_s}{2N_{bin}} \longrightarrow N_{bin} \ge \frac{f_s}{2f_{bin}} 
\end{equation}
Therefore, a compromise between $T_{bin}$ and $f_{bin}$ exists (see \textit{\autoref{eq:_Tbin}} and \textit{\autoref{eq:f_res}}). Thus, when making a spectrogram, \textit{\autoref{eq:f_valid}} must be considered that there is a frequency, $f_{valid}$, from which the transformation is valid \cite{Lara2016Thesis}. 
\begin{equation}\label{eq:f_valid}
	f_{valid}\gtrsim \frac{2f_s}{N_{bin}}
\end{equation}

In acoustics, they are often used to visualize signals that in the time domain are masked by the background noise itself or below another recorded signal with a low Signal-to-Noise Ratio (SNR). 
So a spectrogram becomes a good tool able to study signals with a specific frequency and time duration, looking at the PSD value in the specific bins. This also makes it possible to study the noise level, or the Sound Pressure Level (SPL), received in certain frequency bands \cite{Lara2016Thesis}. If a threshold level is applied to these values, a spectrogram detector can be developed.

Acoustic neutrino detection is not easy because it is a very short signal (high temporal resolution), with a large spectral component (many other signals contained in this bandwidth), and of very low energy (which entails very small amplitudes). One of the characteristics of spectrogram detection is that it does not depend as much on the waveform as on the energy concentrated at certain frequencies during a specific time duration. This presents some advantages over the classical correlation method where the shape of the signal must be known with good accuracy.

\subsection{Trigger alert proposal}
\begin{figure}[htbp]
		\centering
		\includegraphics[width=13cm]{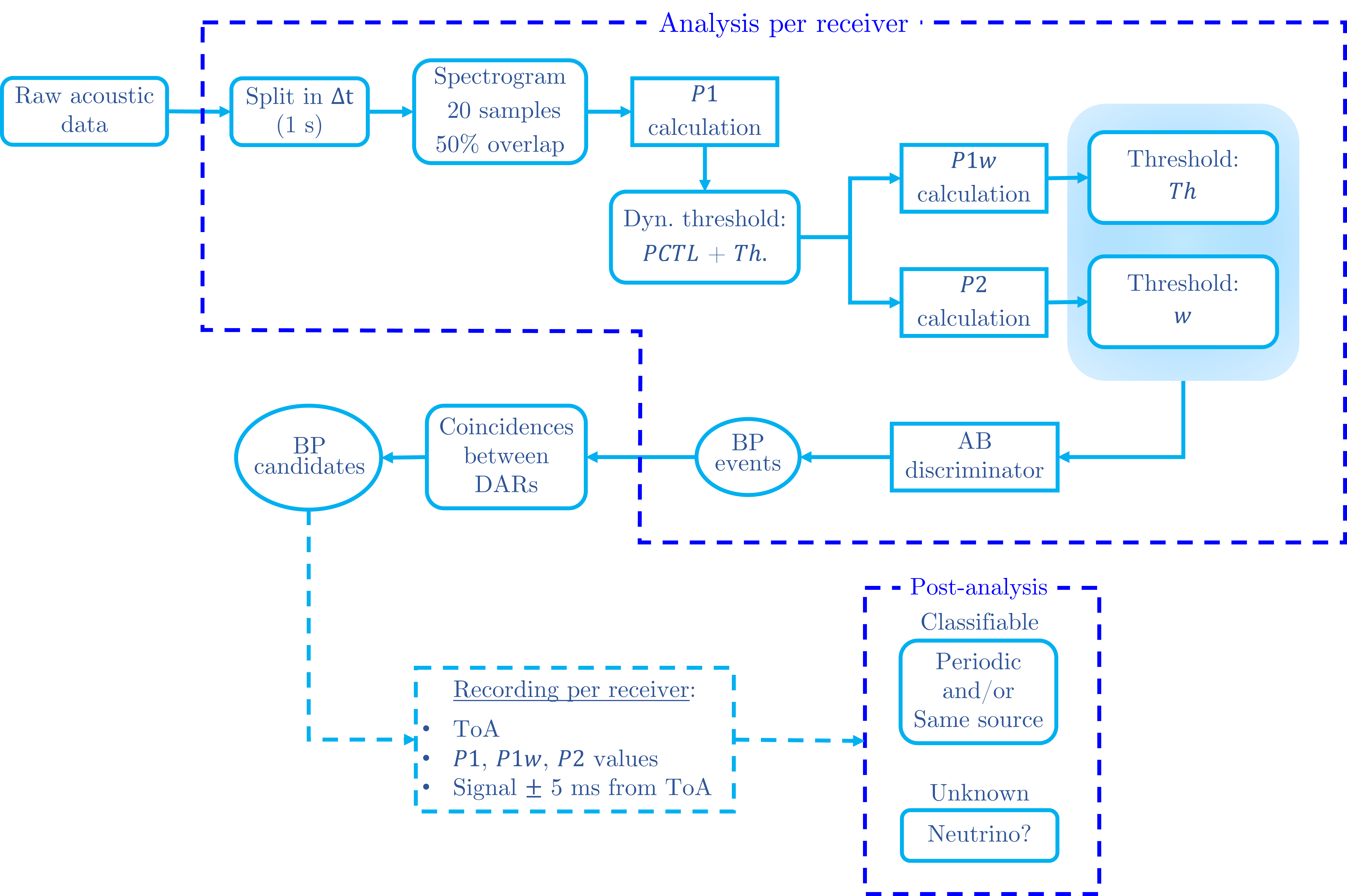}
		\caption{Workflow of the first proposal BP trigger for KM3NeT.}
		\label{fig:TriggerWorkflow}
	\end{figure}

The spectrogram will be used for the analysis of the signals recorded by the acoustic receiver sensors. Using it, the average PSD contained in the bins the BP is expected (20-60 kHz) will be calculated, a parameter called $P1$. If this level exceeds a dynamic threshold that will be set in the spectrogram according to the noise level received by each receiver ($cut1$), two other parameters will be calculated to study the duration of the candidate and cut long signals: $P1w$ ($cut2$) that measures if the high-level is maintained continuously and $P2$ ($cut3$) that takes into account the average level in a wider region ($\pm$50 bins). If a bin of the spectrogram pass the three $cuts$, and does not coincide with any signal emitted by the Acoustic Beacons (AB)\footnote{An AB is the emitter used by the APS. Any candidate that coincides with this signal is disregarded, given that such a powerful and close signal can disturb the spectrum and give a false candidate}, it shall be considered a $BP\ event-seed$. This analysis will be performed for the raw acoustic data recorded by each receptor, if the $BP\ events-seed$ of these are coincident within a time window calculated according to the distance between them, they will be referred to as $BP\ candidate$ (\textit{\autoref{fig:TriggerWorkflow}}).  

Taking into account the sampling frequency of KM3NeT and the BP to be detected, a spectrogram with $N_{bin} =$ 10 and 50\% overlap is proposed, which implies a temporal resolution $T_{bin}$ of about 50 $\mu$s, a frequency resolution $f_{bin}$ of 5 kHz and a valid frequency $f_{valid}$ from 20 kHz upwards. 
In order to maintain common memory levels in KM3NeT, the goal of this first algorithm is to have a candidate rate of less than one per second. 
\section{Testing experimental data}
To perform a first experimental test of the BP detection algorithm, thirteen RUNs (data logging period in KM3NeT) recorded by three ORCA hydrophones have been used (approximately three days of continuous data). This study starts by using hydrophones and not piezoceramic sensors because of their higher sensitivity and bandwidth, flat response, and their fixed position at the base of each DU. 
\subsection{Evaluation}
	Adding an artificial BP to every minute ($\sim$0.02 ev/s) of data will allow to evaluate the behavior of the detector. On the one hand, its accuracy ($Precision$, \textit{\autoref{eq:Precision}}) can be calculated, and on the other hand, its efficiency ($Recall$, \textit{\autoref{eq:Recall}}) can be calculated. Considering a distance of 1 km from the midpoint of the three hydrophones, a point will be randomly selected, which will simulate the position where a Ultra-High-Energy neutrino interacts. Then the Time of Arrival (ToA) and amplitude are calculated for each hydrophone corresponding to that event, and finally, the artificial BP is added to the registered signals before they are analyzed by the detector. 
	When a detected event corresponds to an artificial BP, it will be a True Positive ($TP$). If the candidate is not an artificially added BP, it will be considered as False Positive ($FP$). In addition, a BP added and not detected shall be a False Negative ($FN$). Thus,
		\begin{equation}\label{eq:Precision}
			Precision = \frac{TP}{TP+FP}
		\end{equation}
		\begin{equation}\label{eq:Recall}
			Recall = \frac{TP}{TP+FN}
		\end{equation} 

\subsection{Results}
A previous spectrogram analysis is used to calculate the background noise level (SPL) concentrated in the octave thirds where we look for the BP. \textit{\autoref{tab:SPLtab}} shows the first three quartiles and 99th percentile ($Q25\%$, $Q50\%$, $Q75\%$, and $PCTL99\%$) values for an hydrophone. 
    \begin{table}[htbp]
    \centering
		\begin{tabular}{ccccccc}
    		\hline
    		\multirow{2}{*}{RUN} & \multirow{2}{*}{Date + 6H} & \multicolumn{5}{c}{SPL$_{22-56}$ kHz {[}dB re 1$\mu$Pa{]}} \\
    		                     &                            & $Q25\%$      & $Q50\%$      & $Q75\%$     & $PCTL99\%$ & $(Q75\%-Q25\%)/2$   \\ \hline
    		8008                 & 14/05/2020 00:01                & 64.4         & 64.8         & 65.5        & 80.4    & 0.6       \\
    		8015                 & 15/05/2020 06:01                & 64.0         & 64.0         & 64.1        & 80.2   & 0.1        \\
    		8018                 & 16/05/2020 00:01                & 64.3         & 65.1         & 68.0        & 80.4    & 1.8       \\
    		8019                 & 16/05/2020 06:01                & 64.0         & 64.0         & 64.1        & 79.9    & 0.1       \\
    		8021                 & 16/05/2020 18:01                & 64.9         & 65.5         & 66.0        & 80.4    & 0.6       \\
    		8027                 & 18/05/2020 06:01                & 63.7         & 63.7         & 63.8        & 80.3   & 0.1        \\
    		8042                 & 19/05/2020 18:01                & 64.3         & 64.7         & 65.4        & 80.0   & 0.5        \\
    		8048                 & 21/05/2020 06:01                & 63.7         & 63.7         & 63.8        & 80.1    & 0.1       \\
    		9901                 & 05/05/2021 00:00                & 64.2         & 64.7         & 65.4        & 79.5   & 0.6        \\
    		9907                 & 06/05/2021 12:00                & 63.7         & 64.0         & 64.3        & 79.4   & 0.3        \\
    		9913                 & 08/05/2021 00:00                & 64.1         & 64.2         & 64.3        & 79.5    & 0.1       \\
    		9919                 & 09/05/2021 12:00                & 64.1         & 64.2         & 64.3        & 79.5  & 0.1         \\ \hline
		\end{tabular}
		\caption{Total SPL for the frequencies between 22.3 kHz to 56.3 kHz on the recorded data by the hydrophone in DU3. The semi-interquartile range is presented too.}
		\label{tab:SPLtab}
	\end{table}
	
\begin{figure}[htbp]
		\centering
		\includegraphics[height=4.5cm]{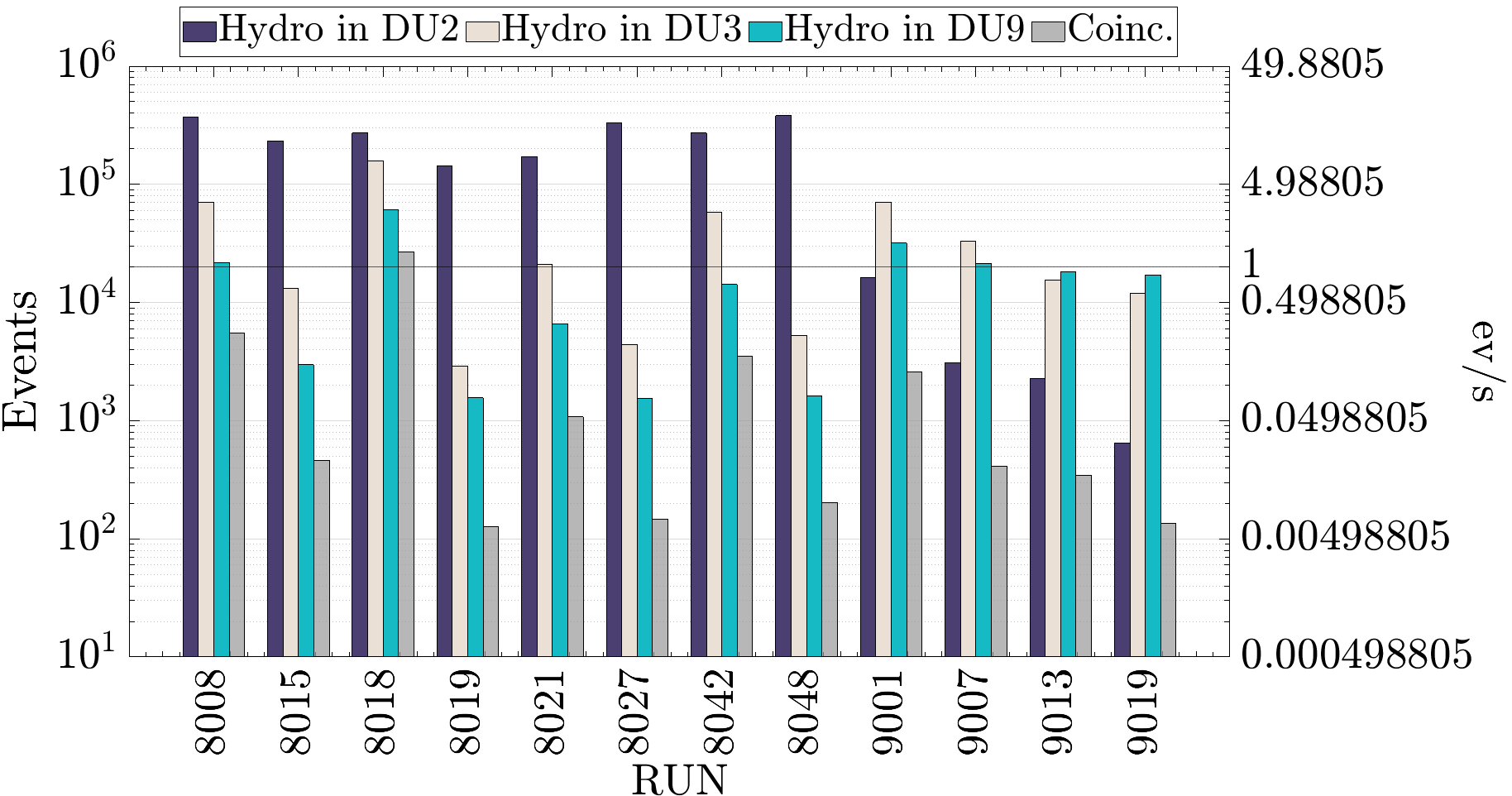}
		\caption{Number of BP events and BP candidates in ev/s for the studied RUNs.}
		\label{fig:ResHist}
	\end{figure}

	\textit{\autoref{fig:ResHist}} shows the number of $BP\ events-seed$ and $BP\ candidates$ and their ratio per second in each hydrophone per physics run. 
	
\begin{figure}[htbp]
		\centering
		\subfloat[]{\includegraphics[width=4.5cm]{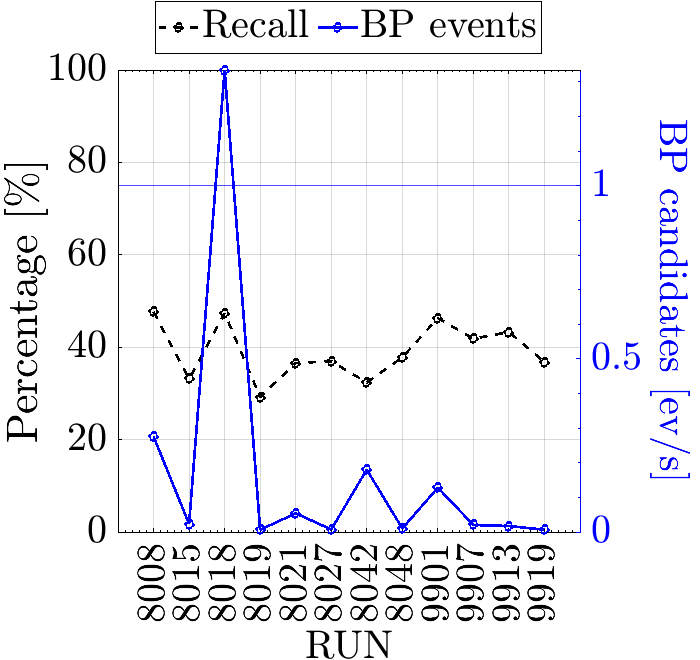}}
		\hspace{.5cm}
		\subfloat[]{\includegraphics[width=4.5cm]{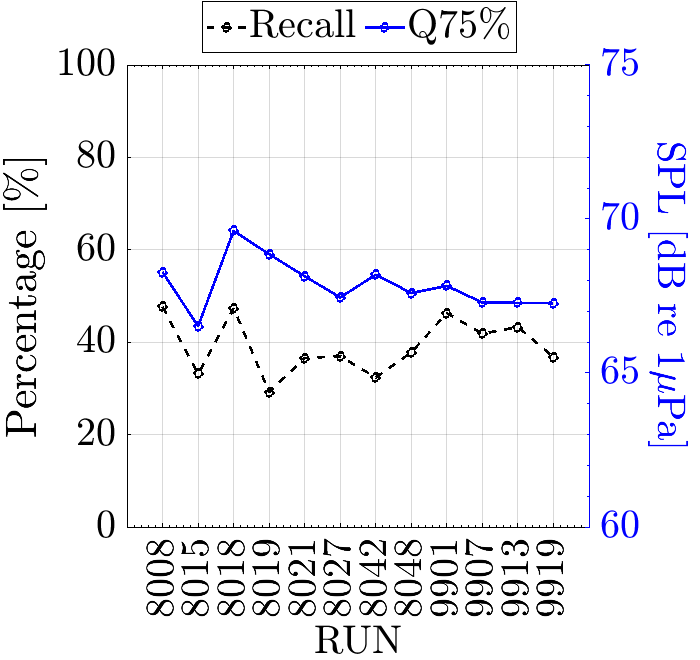}}
		\hspace{.5cm}
		\subfloat[]{\includegraphics[width=4.5cm]{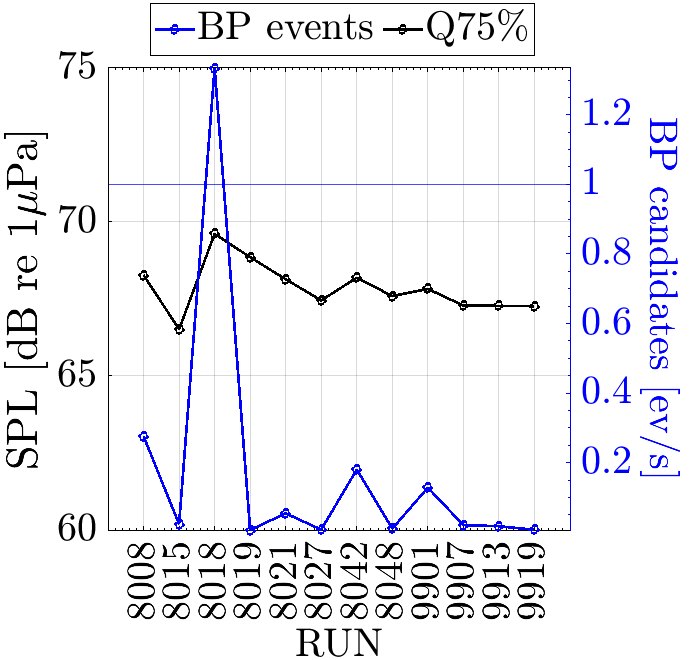}}
		\caption{\textbf{(a)} Efficiency vs Background. \textbf{(b)} Efficiency vs Noise. \textbf{(c)} Background vs Noise}
		\label{fig:ResGraph}
	\end{figure}

	On the other hand, the efficiency ($Recall$), the background (ratio of $BP\ candidates$ per second) and the noise (SPL in quartile 75\%) per RUN are depicted in \textit{\autoref{fig:ResGraph}} to see if there is any trend or correlation between them. From this figure we can check that it is possible to retain $\sim$50 \% of simulated events with the reference signal having a moderate background ($\leq$1 event/s).
	
\begin{figure}[htbp]
		\centering
		\includegraphics[height=5cm]{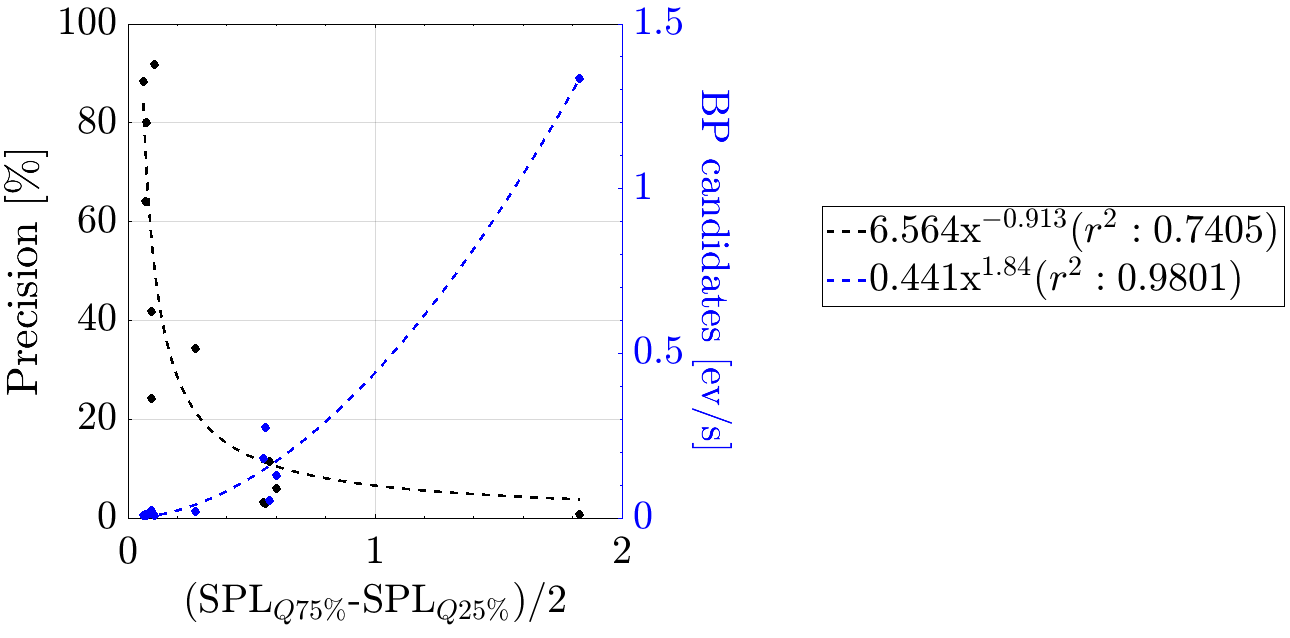}
		\caption{Correlation between the precision and background value (ev/s) in the three hydrophones versus the semi-interquartile range of noise by hydrophone in DU3. Results from fitting the data to potential functions are also shown.}
		\label{fig:Coinc}
	\end{figure}

	Looking at the SPL in the 75\% quartile and semi-interquartile range in \textit{\autoref{tab:SPLtab}}, the event rate per second in \textit{\autoref{fig:ResHist}} and the trends in \textit{\autoref{fig:ResGraph}}, it can be seen that there is some correlation between the noise level and the event rate and the accuracy ($Precision$). \textit{\autoref{fig:Coinc}} shows the effect of noise versus the number of events and the accuracy of the detection algorithm with logarithmic settings. 
	
\section{Conclusions and future steps}
A proposal of trigger for a BP detector, based on the spectrogram, is proposed. The trigger is applicable to the current KM3NeT configuration. The proposal method presents good efficiency, around 50\%, having an acceptable background detection rate, below one event per second, which is assumable and convenient from the point of view of data storage and for further post-processing. 

Now, the idea is to implement the trigger on KM3NeT data anlaysis for each receiver (pizoceramic sensors included), although recommended using quiet data (see \textit{\autoref{fig:Coinc}}). If a $BP\ event$ appears in a lot of receivers, it is possible to estimate if its directivity is narrow or not. In case of wide angle directivity, the candidate should be discarded. The trigger methodology can also be useful to apply searches for other signals, for example for bioacoustics.

\section*{Acknowledgements}
We acknowledge Ministerio de Ciencia e Innovación: Programa Estatal para Impulsar la Investigación Científico-Técnica y su Transferencia (refs. PGC2018-096663-B-C43 and PID2021-124591NB-C42) (MCIU/FEDER), Programa de Planes Complementarios I+D+I (ref. ASFAE/2022/014)  (Generalitat Valenciana/MCI/EU-NextGeneration), Spain.



\end{document}